\documentstyle[12pt,epsfig]{article}
\begin{document}

\begin{center}
{\bf THOMSON SCATTERING OF COHERENT DIFFRACTION RADIATION BY
AN ELECTRON BUNCH.}
\end{center}
\vspace{3mm}
\begin{center}
{ A.P.POTYLITSYN }\\
\vspace{3mm}
{\small \it Tomsk Polytechic University,}\\
{\small \it pr. Lenina 2A, Tomsk, 634050, Russia}\\
{\small \it e-mail: pap@phtd.tpu.edu.ru}
\end{center}
\begin{abstract}
{\small The paper considers the process of Thomson scattering of coherent
diffraction radiation (CDR) produced  by the preceding bunch of the
accelerator on one of the following bunches. It is shown that the yield of
scattered hard photons is proportional to N$_e^3$, where N$_e$ is the number
of electrons per bunch. A geometry is chosen for the CDR generation and an
expression is obtained for the scattered photon spectrum with regard to the
geometry used, that depends in an explicit form on the bunch size.
A technique is proposed for measuring the bunch length using scattered
radiation characteristics. }
\end{abstract}
\vspace{6mm}
\vspace{6mm}

\begin{center}
{ \bf 1. Introduction} \vspace{2mm}
\end{center}
   The process of Compton backscattering (CBS) of the infra-red or visible
 photons by the relativistic electrons was used widely for obtaining
X-ray - and $\gamma$ -- beams with the energy  from $\sim 10^6$eV up to
$\sim 10^{10}$eV [1-4].

The development of laser technologies within recent years has brought up a
 suggestion to use the CBS process for electron bunch diagnostics [5-7].
The authors of an experiment [7] used a femtosecond near infrared terawatt
 laser as a source of radiation which was scattered on a bunch of electrons
with the energy E = 50 MeV. They proposed to use this process for the
measurement of electron bunch characteristics (longitudinal and transverse
bunch sizes, divergence, etc.) The longitudinal bunch structure,
for instance, was measured via the dependence of the scattered hard photon
yield on the time delay between the electron and photon bunches.

It is clear that the accuracy of such measurements relies on the
reproducibility and controllability of characteristics of a powerful laser,
which is a rather complicated problem.

In further works [8,9] it was proposed to measure the bunch length
through such characteristics of coherent transition radiation (i.e. the
transition radiation with a wavelength comparable with the bunch length),
as the radiation spectrum and the autocorrelation function. In the
latter cases one is free from the errors associated with the laser.
However, the methods so far proposed are not  non-destructive (viz.
the  electron beam crosses the foil target).

This paper considers a possibility of electron beam diagnostic using
Thomson scattering of CDR from the preceding bunch on the
following one. Diffraction radiation is produced when a charged particle
moves close to a conducting target. The effects of the target on beam
characteristics could be reduced to an acceptable level by choise of
the distance between the beam and target. Thus, the method proposed here
 is nondestructive as are the methods involving the use of laser emission,
nonetheless, characteristics of the scattered hard radiation are determined
only by the  electron beam parameters.

\begin{center}
{ \bf 2. Thomson scattering of radiation by a moving bunch.} \vspace{2mm}
\end{center}

   During the interaction of an incident photon with a moving electron the
scattered photon energy is to be derived using the conservation laws:

\begin{equation}
\omega_2 = \omega_1\frac{1-\beta\cos\theta_1}{1-\beta\cos\theta_2
+\frac{\displaystyle\omega_1}{E}\{1-\cos(\theta_1-\theta_2)\}} \;.
\end{equation}

Here $\omega_1, \omega_2$ and E are the energies of the incident
and scattered photons and that of the electron, respectively,
$\beta = v/c, v $ is the electron velocity, the angles between the
electron momentum and the incident and scattered photons
$\theta_1,\; \theta_2$ are the same as in [6].
If the primary photon energy and that of the electron satisfy the
conditions
 \begin{equation}
\gamma = E /mc^2 >>1, \ \ \gamma\omega_1 << mc^2\;,
\end{equation}
 the scattering photon energy (1) is linearly dependent on that
of the incident photon :
\begin{equation}
\omega_2 = \omega_1 \frac{1 - \beta\cos\theta_1}{1 - \beta\cos\theta_2}
\approx \omega_1 \frac{2\gamma^2(1-\beta\cos\theta_1)}{1+(\gamma\theta_2)^2}
 \;, 
\end{equation}
where the outgoing photon angle $\theta_2 \sim \gamma^{-1}$.

In a frame where the electron is at rest (ERF), the energy of the photon
scattered, is, according to (2), sufficiently
less than the electron mass.
The photon scattering then occurs virtually without any frequency
changing and, therefore, the scattering process may be described in
terms of classical electrodynamics (Thomson scattering).

In the ERF the classical cross section of  scattering of an electromagnetic
 wave by a free charge [10] is not controlled by its frequency and
is given by  the expression:
\begin{equation}
\frac{d\sigma}{d\rm\Omega'} = \frac{r_0^2}{2}\Big(1+\cos^2\theta'\Big)\;.
\end{equation}
In (5), $r_o = 2.82 \cdot10^{-13}$ cm is the classical radius of an electron,
 and the primes denote the angles  in the ERF.
Transforming these to the laboratory system, we have:
\begin{equation}
\cos \theta'=\frac{\cos\theta_2 - \beta}{1-\beta \cos \theta_2}\;\;,
\end{equation}
\begin{equation}
d\rm\Omega'=\frac{1- \beta^2}{(1-\beta \cos \theta_2)^2} d\rm\Omega
\end{equation}

 From (5) and (6) we obtain the classical  cross section
for the ultrarelativistic case:
\begin{equation}
\frac{d\sigma}{d\rm\Omega} = 4r_0^2 \gamma^2 \frac{1+(\gamma \theta_2)^4}
{[1+(\gamma \theta_2)^2]^4}\;.
\end{equation}

The total cross section derived through integrating  expression (7)
 with respect to angles is the Thomson  cross section:
\begin{equation}
\sigma_{T}=\frac{8}{3} \pi r_0^2\;.
\end{equation}
The yield of secondary photons upon scattering, e.g. of  incident  laser
photons, on a moving electron bunch is to be determined not only by the cross
section of  the process but also by the overlapping of the laser and
electron  beams in space and time, which is characterized by luminosity $L$:
\begin{equation}
\frac{d N_2}{dt}=L \sigma_T \;. 
\end{equation}
Let  us consider the head-on collision of electron and photon bunches.
Luminosity in this case is defined as follows:
\begin{equation}
L = c N_{e} N_{ph} F \int \int \int \int dx dy dz dt f_{ph}(x,y,z+ct)
f_{e}(x,y,z-\beta ct)\;.
\end{equation}

Here $N_e\;,\; N_{ph}$  are the number of particles in the electron and
photon bunches, $f_e\;,\; f_{ph}$  are the corresponding normalized
electron and photon distributions and F is the collision frequency of
the bunches. For the  monodirected beams with a Gaussian
distribution in both transversal and longitudinal directions:
\begin{eqnarray}
f_e &=& \frac{2}{(2 \pi)^{3/2} \sigma_e^2 l_e} \exp\biggl\{-\frac{r^2}
{\sigma_e^2}
-\frac{(z-\beta ct)^2}{2 l_e^2}\biggr\}\;,\nonumber \\
f_{ph} &=& \frac{2}{(2 \pi)^{3/2} \sigma_{ph}^2 l_{ph}}
 \exp\biggl\{-\frac{r^2}{\sigma_{ph}^2}- \frac{(z+ct)^2}{2 l_{ph}^2}
\biggr\}\;, \\
r^2 &=& x^2 + y^2\;,\nonumber  
\end{eqnarray}
the luminosity is readily calculated
 \begin{equation}
 L = N_eN_{ph}F \frac{1}{2\pi (\sigma_e^2 + \sigma_{ph}^2)}\;.
 \end{equation}

In (11), $\sigma_e^2\;,\;\sigma_{ph}^2$ are the variables  characterizing
the transversal and $l_e^2\;,\;l_{ph}^2$ are those for the longitudinal
distributions. For the head--on collisions it follows from (12) that the
luminosity is governed solely by the transverse dimensions of the electron
 and photon bunches. The number of the photons scattered through
collision of single bunches can be estimated from (9) and (12):
\begin{equation}
N_2= \frac{1}{2}N_{ph} \frac{N_e \sigma_{T}}{S_e + S_{ph}}\;,
\end{equation}

where $S_e, S_{ph}$ are the cross--sections of the electron and photon
 bunches. The value $S = \displaystyle\frac{1}{2}N_e\displaystyle
\frac{\sigma_{T}}{S_e+S_{ph}}$ can be treated as the reflectivity of
the electron bunch. For the electron numbers and bunch size attainable
this value is considerably small.
Therefore, one typically uses radiation of a powerful laser as a primary beam.

 However, effective overlapping of the laser and accelerator bunches  is a
difficult task, while linear dependence of the scattered beam intensity (8)
on the number of electrons in the bunch poses natural restrictions on the
intensity of the resulting X-ray or $\gamma$-beam.
If a beam of incident photons  is to be generated by one of the preceding
 electron bunches of the accelerator, then the temporal and longitudinal
structures of the colliding bunches will be similar.

In the experiment [11] a incident beam of infra-red radiation ($\lambda =3.5
\div$ 7 mcm) was generated by the electrons with the energy E = 50 MeV
$(\gamma \sim$ 100) in an undulator with $\sim 4$ m length. The electron beam
parameters satisfied the gain mode of the free electron laser.

It seems possible that one can use a beam of coherent radiation
of a short electron bunch as a primary beam of soft photons. In
this case, the radiation intensity in the wavelength region $\lambda_1$,
comparable with the bunch length, is quadratically dependent
on the number of electrons in the bunch [12], which compensates for
the low reflectivity of the electron bunch. Instead of a laser source,
coherent diffraction radiation (CDR), i.e. the radiation produced
while a short bunch of electrons is passing close to a
 metal target [13], can be taken as a source of primary radiation.

Fig. 1 shows a potential experimental scheme. Electron bunches pass through
a circular opening of the radius R in a metal target,
which results in generation of CDR in the wavelength region
$\lambda_1\geq l_e$, the electrons are deflected
by a  bending magnet BM, while  CDR is reflected and focused  by a
 thin concave mirror CM on one of the following bunches.
The scattered photons  with the energy corresponding to the X-ray
region  are extracted through the  center hole of the mirror CM,
suffering but a  small loss.
The distance between  the center hole of the mirror and the target,
$L_o$, is selected from the condition
\begin{equation}
2 L_o=\frac{L_B}{\beta } \cdot m\;\;,\;\;\;m=1,2,3... \;, 
\end{equation}

where $L_B$ is the distance between the bunches.

The spectrum of the photons backscattered by a single electron may be
calculated in the following manner:
\begin{equation}
\frac{dN_2^0}{d\omega_2} = const\int\int d\rm\Omega_2d\omega_1
\displaystyle\frac{dN_1}{d\omega_1 }
\  \displaystyle\frac{d\sigma}{d\rm\Omega_2}\delta\Bigl(\omega_2-\omega_1
\displaystyle\frac{4\gamma^2}{1+(\gamma\theta_2)^2}\Bigr)\;.
\end{equation}

Here $\displaystyle\frac{dN_1}{d\omega_1}$ is the spectrum of the
incident photon beam.
Integration in (15) should be carried out with respect to  all
the spectral region of the initial radiation and the exit aperture
 $\Delta\rm\Omega_2$.

The yield of photons scattered by an electron bunch is described by a more
complicated formula:
\begin{equation}
\frac{dN_2^{B}}{d\omega_2}=\int\int d\rm\Omega_2d\omega_1\frac{dN_1}
{d\omega_1} \frac{d\sigma}{d\rm\Omega_2} \ \frac{N_e}
{2\pi(\sigma_e^2+\sigma_{ph}^2)}
\delta\Big(\omega_2-\omega_1\frac{4\gamma^2}{1+(\gamma\theta_2)^2}\Big)\;. 
\end{equation}

\begin{center} {\bf 3. Spectrum of coherent diffraction radiation.}
  \vspace{2mm}
\end{center}

DR spectrum may be calculated numerically using the results of work [14]
for the spectral--angular density of the energy radiated  from a single
electron passing through a circular opening with the radius  R in an ideal
conductor:
\begin{equation}
\frac{d^2W}{dxd\rm\Omega}= \frac{\alpha\omega_c}{\pi^2} \frac{\sin^2\theta}
{(\sin^2\theta+\gamma^{-2})^2} J_0^2\Bigl(\frac{x}{2}\gamma\sin\theta\Bigr)
K_1^2\Bigl(\frac{x}{2}\Bigr) \Bigl(\frac{x}{2}\Bigr)\;,  
\end{equation}
where $\alpha$ is the fine structure constant, $\displaystyle\omega_c =
 \frac{\gamma}{2R}$ is the characteristic energy of DR, $\theta$--outgoing
photon angle,   $\omega_1$ \
is the  energy of emitted photon, $x = \omega_1/\omega_c$
is the  dimensionless energy variable.
From here up to the end of paper there will be used the system of units
$\hbar = m =c = 1$.

In  expression (17) $J_0(x)$ is the Bessel function of the zeroth order,
$K_1(x)$ is the  modified Bessel function.
 From (17) one may obtain the DR intensity spectrum
$\displaystyle\frac{dW}{dx}$
after integration with respect to the  solid angle covered by the reflected
mirror.
Calculated spectra for apex angles $\theta_{1m} = k_1/\gamma \ (k_1 = 5,10)$
are shown in Fig. 2.

Following [12] one may write the spectrum of CDR emitted by
a bunch of  N$_e$ electrons  as below:
\begin{equation}
\frac{dN_1^B}{d\omega_1}= N_e(1+ f(\lambda_1)N_e)\frac{dN^0}{d\omega_1}
\approx N_e^2 f(\lambda_1)\frac{dN^0}{d\omega_1}, \ \lambda_1 \geq l_e\;.
\end{equation}

Here $\lambda_1 $ is the wavelength of DR and  $f(\lambda_1)$ is
the bunch form factor defined as the squared Fourier transform
of longitudinal distribution of electron density in a bunch.
For the Gaussian distribution (11) we have:
\begin{equation}
f(\lambda_1) = \Big| \frac{1}{\sqrt{2\pi} l_e}\int
\exp\biggl\{ - \frac{z^2}{2 l_e^2}\biggl\}
\exp\biggl(-i\frac{2 \pi z}{\lambda_1}\biggr)dz\Big|^2
= \exp\biggl(-\frac{4\pi ^2 l_e^2}{\lambda_1^2}\biggr)
= \exp\big(-\omega_1^2l_e^2\big)\;.
\end{equation}

The photon DR spectrum may be easily derived from the DR intensity spectrum:
\begin{equation}
\frac{dN^0}{d\omega_1} = \frac{1}{\omega_1}\cdot \frac{dW}{d\omega_1}=
\frac{1}{\omega_1}\cdot\frac{dW}{\omega_c dx}\;.
\end{equation}

It is clear that there are two energies characterizing the spectrum (18):
\begin{equation}
\omega_{ch1}\sim\omega_c= \frac{\gamma}{2R}\;, \ \omega_{ch2} \sim
\frac{1}{l_e},
\end{equation}
one of them $\omega_{ch1}$ connected with the DR spectrum from a single
electron and  the other ($\omega_{ch 2}$)-- with the collective emission
from the bunch.

For an ultrarelativistic electron beam the transversal and longitudinal
sizes of a bunch may be less than 1 mm. In a similar case, one may use
a  hole with the radius R about a few millimeters. So, we may consider the
case
\begin{equation}
\frac{\gamma}{2R} \gg \frac{1}{l_e}\;.
\end{equation}

It means that the coherent effects are significant in the region
$\omega_1 \ll \omega_c$ where  the DR intensity spectrum may be taken
as a constant (see Fig. 2).
In the limit $\omega_1 \rightarrow 0(x \rightarrow 0)$ we have

\begin{equation}
\frac{dW}{d\omega_1} \approx \frac{\alpha}{\pi}
\biggl\{ln(1+k_1^2)-\frac{k_1^2}{1+k_1^2}\biggr\} =
 \frac{\alpha}{\pi}C_{\parallel}\;. 
\end{equation}

After all substitutions one may obtain:
$$
\frac{dN_2^B}{d\omega_2} = \frac{2}{\pi^2}\alpha r_0^2N_e^3C_{\parallel}
\int\int d\omega_1d {\rm\Omega}_2 \ \displaystyle\frac{1}{\omega_1}
 \displaystyle\frac{\gamma^2[1+(\gamma\theta_2)^4]}
{[1+(\gamma\theta_2)^2]^4}\displaystyle\frac{\exp
(-\omega_1^2 l_e^2)}
{(\sigma_e^2 + \sigma_{ph}^2)}\times  \nonumber  $$
$$
\times\delta\Big(\omega_2 - \omega_1\displaystyle\frac{4\gamma^2}
{1+\big(\gamma\theta_2\big)^2}\Big)\;.\eqno(24)
$$

In  formula (24) the denominator has the value $\sigma_{ph}^2$
characterizing the radius of the focused photon beam in the interaction
point. Due to the diffraction limit the size of  the light spot cannot be
less than $\lambda_1/2\pi$. So, for estimations we shall use the latter
value instead of $\sigma_{ph}$.

As one may see from (24) the scattered yield has the cubic dependence on
 the number of electrons per bunch.

Other authors [15,16] considered electromagnetic radiation produced by the
 collision of short electron bunches and also arrived at a cubic
dependence of the photon yield with the energy  $\omega < \displaystyle
\frac{4\gamma^2} {l_e}$  during collision of identical bunches.

Roughly speaking, the works mentioned earlier studied scattering of the
field of virtual photons of one bunch on the other, while
this paper deals with the process where real photons emitted by the
preceding bunch are scattered on one of the following bunches.

\begin{center}
{\bf 4. Dependence of characteristics of scattered photons \\
on electron bunch parameters. } \vspace{2mm}
\end{center}

Due to narrow angular distribution of the scattered photons decreasing as
($\gamma\theta_2)^{-4}$ for large emission angle $\theta_2 \gg \gamma^{-1}$
 eq.(24) may be simplified:
$$
\frac{dN_2^B}{d\omega_2}= \frac{2}{\pi^2} \alpha r_0^2 N_e^3\gamma^2
\Delta\Omega_2C_{\parallel}\int d\omega_1\delta(\omega_2 - 4\gamma^2\omega_1)
\displaystyle\frac{\exp\big(-\omega_1^2 l_e^2\big)}{\omega_1\Big(\sigma_e^2 +
\displaystyle\frac{1}{\omega_1^2}\Big)}\;,
\eqno(25)
$$
if the exit aperture  $\Delta\Omega_2 = \pi\theta_{2 \rm max}^2$
is comparable with $\gamma^{-2}$:
$$ \theta_{2 \rm max} = k_2\gamma^{-1}, \ k_2 \sim 1.$$

Using the well--known property of the  $\delta$--function we may obtain:
$$
\frac{dN_2}{d\omega_2}= \frac{2}{\pi}\alpha r_0^2 N_e^3 C_{\parallel}k_2^2
\frac{\exp\biggl[\displaystyle -\biggl(\frac{\omega_2l_e}{4\gamma^2}\biggr)^2
\biggr]}
{\omega_2\biggl[\sigma_e^2+\biggl(\displaystyle\frac{4\gamma^2}{\omega_2}
\biggr)^2\biggr]}= \displaystyle\frac{2}{\pi}\alpha r_0^2 N_e^3
C_{\parallel} k_2^2 \displaystyle\frac{F}{\omega_2}  \eqno(26)
$$

One may see from (26) that the yield of scattered photons does not depend
on the electron energy, if the maximum outgoing angle $\theta_{2 \rm max}$ is
measured in units $\gamma^{-1}$. Of course, the scale of transformation
of the photon energy is defined by the electron energy (see Eq.(3)).

The spectrum (26) is shown in Fig. 3 for different ratios between
$\sigma_e $ and $l_e$.
There are the clear broad maxima whose positions are determined by
the ratio $r = \sigma_e / l_e$. With  this ratio decreasing spectral
maximum shifts to the value
$$
\displaystyle \omega_{2\rm max} = \displaystyle\frac{1}{\sqrt{2}}
 \displaystyle\frac{4\gamma^2}{l_e}
$$
and  the intensity rises due to increasing luminosity.
Let  us estimate the photon yield at the maximum for following parameters:
$N_e=10^{10}e^{-}$/bunch; \  $ \sigma_e=l_e$=1mm; \
$k_1=10(C_{\parallel}=3.6)$; \  $k_2$=3; \ $\Delta\omega_2 /\omega_2
= 10\%$.

Then
$$
\Delta N_2^B = \frac{dN_2^B}{d\omega_2}\Delta\omega_2 = \frac{2\alpha}{\pi}
\biggl(\frac{r_0}{l_e}\biggr)^2 N_e^3C_{\parallel}k_2^2 F_{\rm max}
\frac{\Delta\omega_2}{\omega_2} = 3.7\cdot10^{4} ph/bunch.
$$
For the electron energy E = 1000 MeV the photons scattered at the spectral
maximum have the energy around 1.6 keV.

However, the estimation of the yield obtained above is valid
only if the focusing mirror is located at a large distance from the target.
$$
L_0 \gg L_f\;. \eqno(27)
$$
Here $L_f$  is the formation length  that characterizes the distance
at which  the radiation  of the wavelength $\lambda $, propagating
at the angle $\theta $, is completely separated  from the initiating charge:
$$
L_f=\frac{\beta \lambda}{1-\beta \cos \theta} \eqno(28)
$$
For  forward emission $(\theta_1 \sim \gamma^{-1})$ in the ultrarelativistic
case $(\gamma \geq 10^2)$  the CDR formation length
$$
L_f \approx \frac{2 \gamma^2 \lambda_1}{1+\gamma^2 \theta_1^2}\;. \eqno(29)
$$
can exceed tens of meters. In a real case the mirror CM (Fig. 1)
can be placed at a distance $L_0 \ll L_f$ . Then the  DR intensity
(initial photon flux) is suppressed as $(L_f/L_0)^2$ [17].
For the case considered,  the suppression factor may reach
$\sim 10^{-4} \div 10^{-5}$ for a distance between target and mirror
about a few meters.

As follows from (28), for the emission angles $\theta_1 \sim \pi/2$
the formation length is comparable  with the wavelength. For these
large emission angles the mirror positioned  at $L_0 \gg \lambda_1$
 does not effect the DR intensity.
Fig. 4 shows the schematic of a potential  application of the proposed
geometry. An electron beam passes in the vicinity of a metal target tilted
 at $\theta=45^{\circ}$ with respect to the electron momentum,
CDR propagates at $\theta_1 \approx  90^{\circ}$ to the beam (in a
close analogy with backward transition radiation [18]).

Spectral-angular distribution of DR when a single charge passes near a
 tilted ideally conducting semi-plane was obtained in [19]. For the
ultrarelativistic case, when we introduce the angles $\theta_x$,
 \  $\theta_y$ measured from the direction of mirror
 reflection (the $x$--axis is oriented along the target edge),
the spectral-angular distribution of DR is written in a simpler form [20]:

$$
\frac{d^2W}{d\omega_1 d\Omega} = \frac{\alpha}{4\pi^2}
\exp\biggl(-\frac{\omega_1}{\omega_c}\sqrt{1+\gamma^2\theta_x^2}\biggr)
\frac{\gamma^{-2}+2\theta_x^2}{(\gamma^{-2}+\theta_x^2)
(\gamma^{-2}+ \theta_x^2 +\theta_y^2)} \eqno(30)
$$

Here $\omega_c =\displaystyle \frac{\gamma}{2a}$, \ $a$ is the spacing
 between the particle trajectory and the edge of the target.

Fig. 5 shows the DR intensity spectrum, $ \displaystyle\frac{dW}
{d\omega_1}$, obtained by integrating expression (30) with respect to the
 focusing mirror aperture $ \theta^2 = \theta_x^2 + \theta_y^2
\leq (k_1\gamma^{-1})^2$ for $k_1$ = 5  and 10. In contradiction with
the case of passing through the centre of the hole, the spectrum
$\displaystyle\frac{dW}{d\omega_1}$ in the energy range
$\omega_1 \ll \omega_c$ will be aproximated by a linear dependence:

$$
\frac{dW_1}{d\omega_1} = \frac{\alpha}{\pi}C_{\perp}\Big(1-B(\theta_{1m})
\displaystyle\frac{\omega_1}{\omega_c}\Big)\;,
 \  C_{\perp}= \frac{\alpha}{2\pi}
\biggl\{ln(1+k_1^2) + \frac{1}{\sqrt{1+k_1^2}}-1\biggr\}\;,
 \eqno(31)
$$
where $B(5\gamma^{-1}) \approx 2.6.$

The luminosity for the 90$^{\circ}$ collision of bunches described by
distributions (11) can also be calculated analytically:

$$
L = c N_eN_{ph}F \int\int\int\int dx dy dz dt f_{ph}
(x,y,z + ct) f_e(x,y + \beta ct,z ) =   \nonumber
$$
$$
=\frac{N_eN_{ph}F}{\pi\sqrt{(\sigma_e^2 + \sigma_{ph}^2)
(\sigma_e^2 + \sigma_{ph}^2 + 2l_{ph}^2 + 2l_e^2)}}\;.
\eqno(32)
$$

Using the same approximations as in deriving expression (26)
we can arrive at:

$$
\frac{dN_2^B}{d\omega_2} = \frac{4}{\pi}\alpha r_0^2N_e^3C_{\perp}
k_2^2  \frac{\exp\biggl[-\biggl(\displaystyle\frac{\omega_2l_e}
{2\gamma^2}\biggr)^2 \biggr]}
{\omega_2\sqrt{\biggl[\sigma_e^2 +
\biggl(\displaystyle\frac{2\gamma^2}{\omega_2}
\biggr)^2 \biggr]\biggl[\sigma_e^2 + \biggl(\displaystyle\frac{2\gamma^2}
{\omega_2}\biggr)^2+4l_e^2 \biggr]}} =  \nonumber
$$
$$
 = \frac{4}{\pi} \alpha\biggl(\frac{r_0}{l_e} \biggr)^2
N_e^3C_{\perp}k_2^2
 \frac{ \exp\biggl[-\biggl(\displaystyle\frac{\omega_{2}l_e}
{2\gamma^2}\biggr)^2\biggr]}
{\omega_2\sqrt{\biggl[r^2 +
 \biggl(\displaystyle\frac{ 2\gamma^2}{l_e\omega_2}\biggr)^2\biggr]
 \biggl[r^2 + 4 + \biggl(\displaystyle\frac{2\gamma^2}
{l_e\omega_2}\biggr)^2\biggr]}}\;. \eqno(33)
$$

For the geometry considered the coefficient of  frequency transformation
is twice as small as compared with the head--on collision
(see formula (3)).

Depicted  in Fig. 6 is the scattered photon spectrum calculated following
 formula (33). Similar to the head--on collision the spectrum has a
maximum in  the region of energies
$$
\omega_{2 m}\approx 0.5\cdot\frac{2\gamma^2}{l_e}\;.
$$

Estimation of the scattering photon yield for the geometry considered
here for the same conditions as before gives a close value:

$$
\Delta N_2^B = 2.9 \cdot 10^{4} ph/bunch.
$$

Contrary to the geometry used previously, however, in this case
the radiation forming length coincides with the wavelength
($\lambda_1 \sim 1$mm). Therefore, the focusing mirror positioned
at a distance $\L_0 \gg \lambda_1$ would not cause any suppression of the
DR yield, and the resulting expression (33) could be used for estimation
of the hard photon yield when planning an experiment.

Notewortly is the fact that when calculating the luminosity (32) it was
assumed that the  centers of the photon and electron bunches
pass the interaction point at  the same time. Should the
focusing mirror be placed with a certain error $\Delta L_0$, then there
would appear an additional term in expressions (32),(33):

$$
D(\Delta L_0) = \exp\Bigl\{ -  \frac{\Delta L_0^2}
{\sigma_e^2 + \sigma_{ph}^2 + 2(l_e^2 + l_{ph}^2)}\Bigr\}\;.\eqno(34)
$$
For the frequent case, $\sigma_e < l_e$, one can get the information on
the electron bunch length via measuring the scattered photon yield
versus $\Delta L_0$ (detuning curve), since $l_{ph} = l_e$.

\begin{center}
{\bf 5. Conclusion.} \vspace{2mm}
\end{center}
As discussed above, the energy of scattered photons for the case of
ultrarelativistic  electrons ($\gamma \geq$ 1000) corresponds to the X-ray
region, while for moderate relativistic energies ($\gamma \leq $100)
the secondary photon  spectrum would include the visible range.
It is known that the common techniques for electron beam diagnostics based
on detection of optical transition radiation do not allow us to
measure the length of  submillimeter bunches. In this context,
measurement of the detuning curve by mechanical displacement
of the focusing mirror seems to offer a means for measuring even
shorter bunches with the use of  simpler equipment than a
streak camera.

It should be noted that the CBS process of laser photons on an electron
bunch was considered in 90$^\circ$ geometry [21], and it was shown
that for a certain geometry and bunch parameters the yield of scattered
photons may be by 2-3 orders exceed that from scattering on $N_e$
electrons independent of each other. The enhancement factor, dictated by
the coherent compton scattering, is proportional to $N_e\displaystyle
\frac{\lambda_1}{\gamma^2}$.
It is to be expect that during scattering of CDR on the following electron
bunch the effect of coherence could be made manifest in as more
pronounced fashion, since the wavelength of primary radiation is by
2-3 orders higher than laser emission wavelength and, secondary,
the coherent Thomson scattering would involve the dependence of the
number of secondary photons on the number of electrons per bunch to be
proportional to $N_e^4$.

\begin{center}
{\bf 6. Acknowledgments.}  \vspace{2mm}
\end{center}
The author is grateful to Prof. M.Ikezawa and Dr. Y.Shibata for helpful
 discussions and also appreciates the assistance of T.D.Litvinova,
L.V.Puzyrevich and O.V.Chefonov in preparing the text of the paper
for publication.

\vspace{6mm}
\centerline{\large\bf  References}
\vspace{5mm}
\begin{enumerate}
\item
 O.F.Kulikov, Y.Y.Telnov, E.I.Filippov et al. Phys.Lett. {\bf 13},
344(1964)
\item
C.K.Sinclair, J.J.Murray, P.R.Klein et al. IEEE Trans. Nucl.Sci.
{\bf16}, 1065(1969)
\item
L.Federici, G,Giordano, G.Matone et al. Nuovo Cim. {\bf B59}, 247(1980)
\item
 G.Ya.Kezerashvili, A.M.Milov, B.B. Woitsekhowski. Nucl.Instr. and
 Meth. {\bf A328}, 506(1993)
\item
 T.Shintake. Nucl.Instrum. and Meth. {\bf A311}, 453(1986)
\item
 Ian C.Hsu, Cha-Ching Chu  and Chuan-Ing Yu. Phys.Rev. E{\bf 54},
5657(1996)
\item
 W.P.Leemans, R.W.Schoenlein, P.Volfbeyn et al. Phys.Rev.Lett.
{\bf 77}, 4182(1996)
\item
 Y.Shibata, T.Takahashi, T.Kanai et al. Phys.Rev. E{\bf 50},
1479(1994)
\item
  R.Lai, A.J.Sievers. Phys.Rev. E{\bf 50}, R3342(1994)
\item
J.D.Jackson. Classical Electrodynamics, Wiley, New York, 1975
\item
F.Glotin, J.-M.Ortega, R.Prazeres et al. Phys.Rev.Lett.
{\bf 77}, 3130(1996)
\item
T.Nakazato, M.Oyamada, N.Nimura et al. Phys.Rev.Lett.
{\bf 63}, 1245(1989)
\item
Y.Shibata, S.Hasebe, K.Ishi et al. Phys.Rev. E{\bf 52},
6787(1995)
\item
Yu.N.Dnestrovskii and D.P.Kostomarov. Sov.Phys.Dokl.
{\bf 4}, 158(1959)
\item
M.Bassetti, J.Bosser, M.Gygi-Hanney et al.
IEEE Trans.Nucl.Sci.\\ NS-30, 2182(1983)
\item
I.F.Ginzburg, G.L.Kotkin, S.I.Polityko et al.
JETP Lett. {\bf 55}, 637(1992)
\item
Y.Shibata, K.Ishi, T.Takahashi et al. Phys.Rev. A{\bf 45},
R8340(1992)
\item
L.W.Wartski, S.Roland, J.Lassale et al.
J.Appl.Phys. {\bf 46}, 3644(1975)
\item
A.P.Kazantsev and G.I.Surdutovich. Sov.Phys.Dokl. {\bf 7},
990(1963)
\item
A.P.Potylitsyn. Nucl.Instrum. and Meth. {\bf B145},169(1998)
\item
G.Giordano, G.Matone, A.Luccio et al. Laser and Particle
Beams,\\ {\bf 15}, 167(1997)
\end{enumerate}

\newpage
\begin{figure}
\centering
\unitlength=1cm
\begin{picture}(18,18)
\put(-1,-1){\epsfxsize=18cm\epsfbox{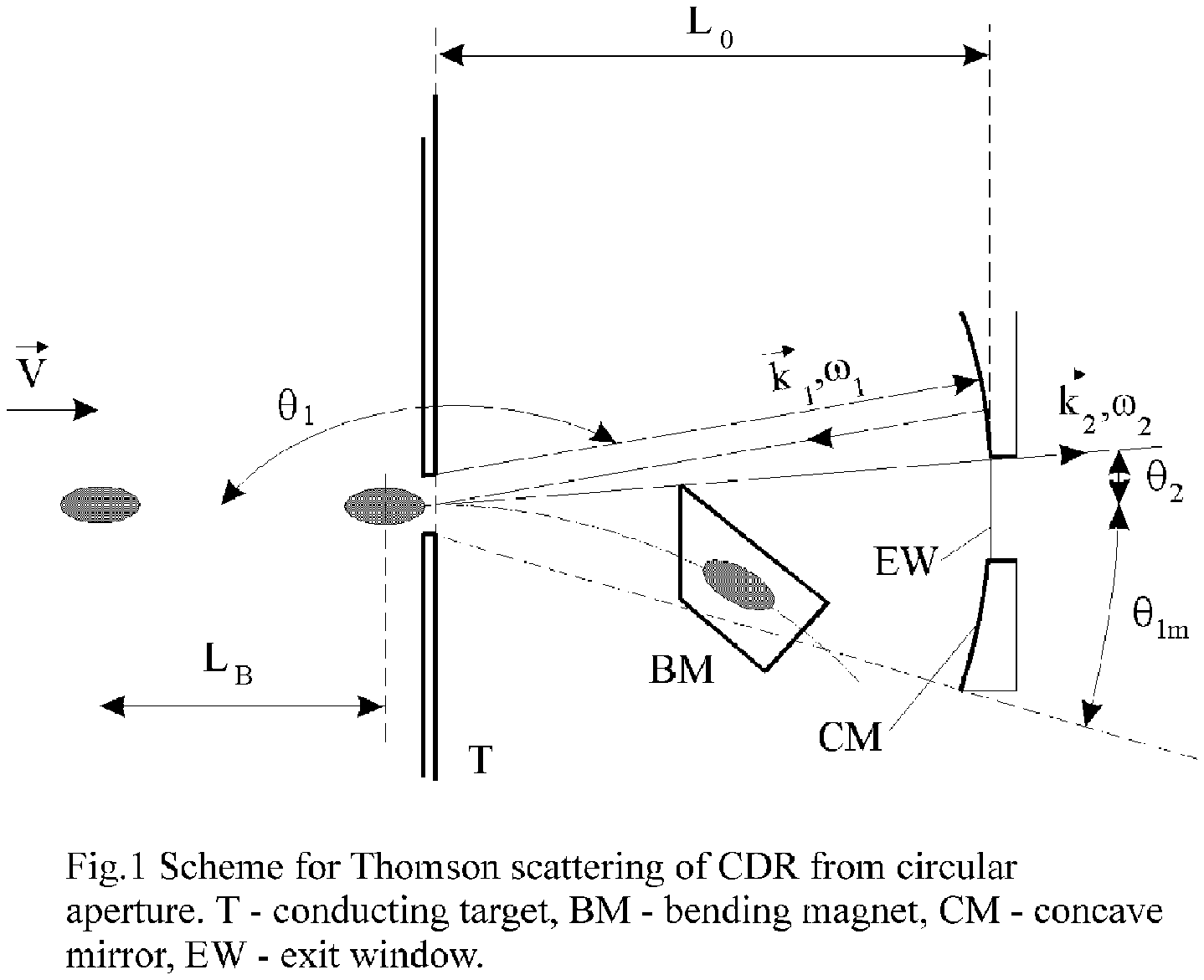}}
\put(-1.8,-4){\epsfxsize=18cm\epsfbox{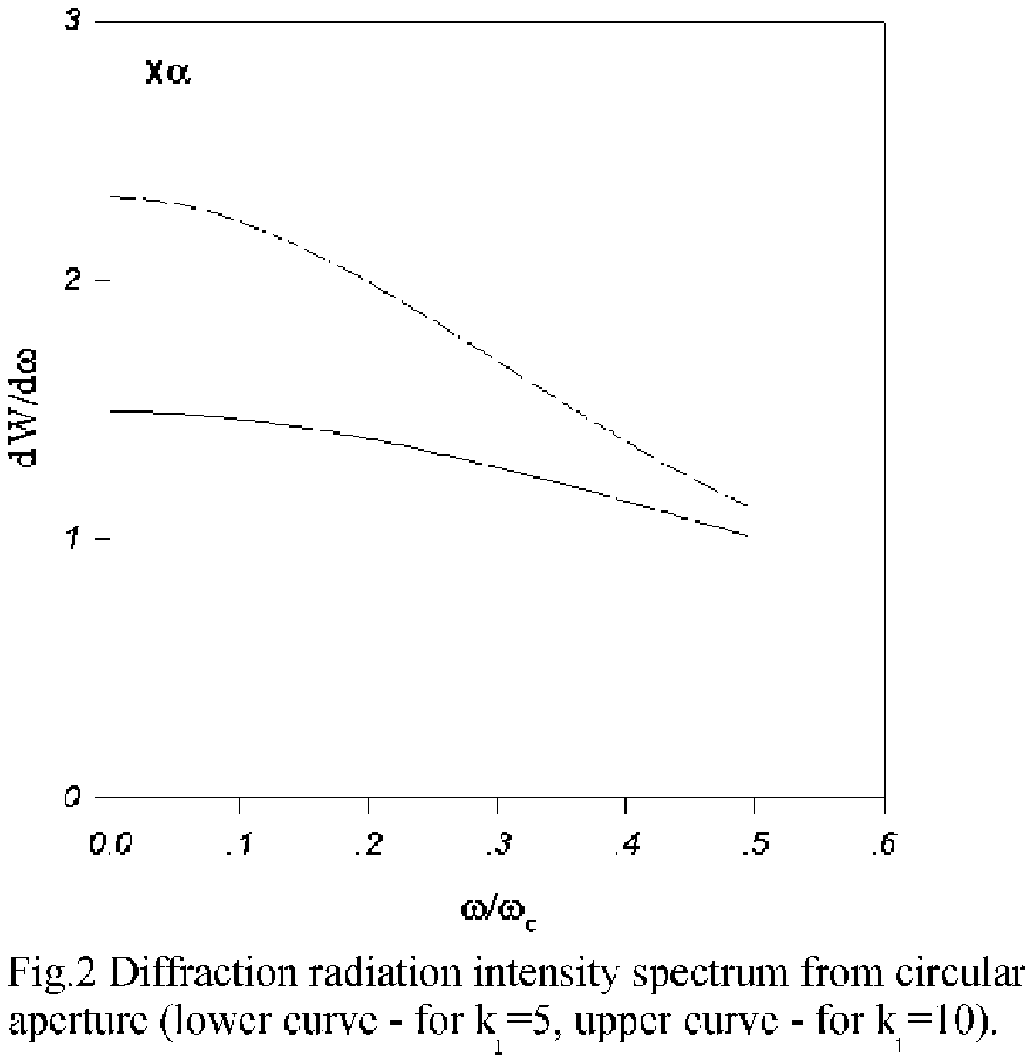}}
\end{picture}
\end{figure}

\newpage
\begin{figure}
\centering
\unitlength=1cm
\begin{picture}(18,18)
\put(-1.2,-2){\epsfxsize=18cm\epsfbox{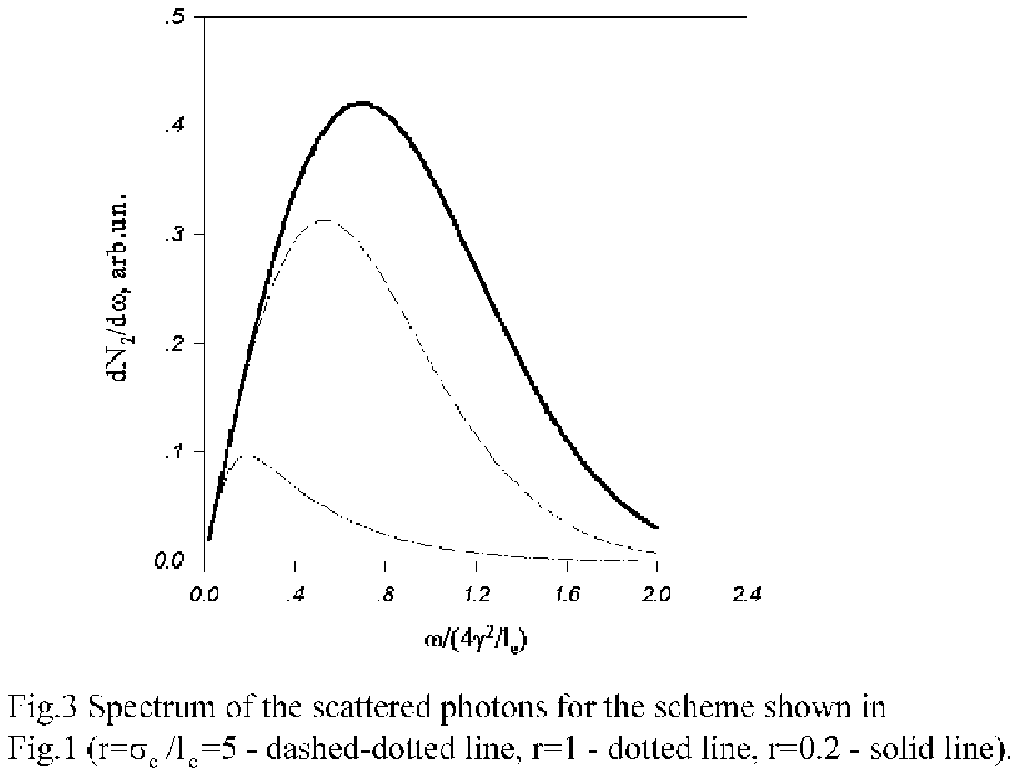}}
\put(-3,-5){\epsfxsize=18cm\epsfbox{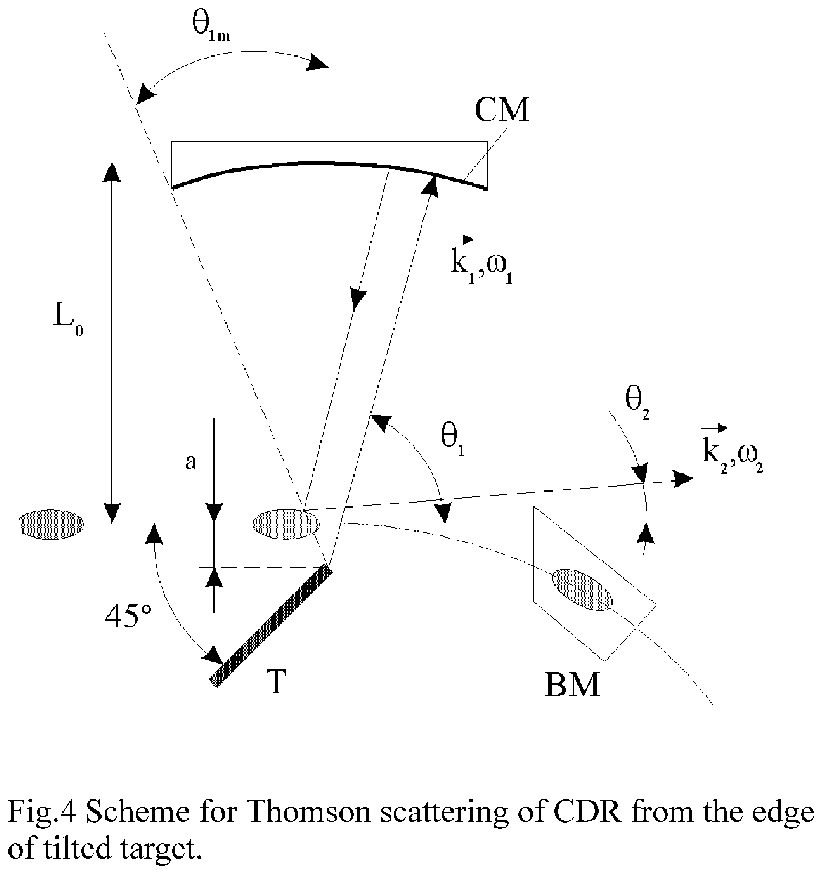}}
\end{picture}
\end{figure}

\newpage
\begin{figure}
\centering
\unitlength=1cm
\begin{picture}(18,18)
\put(-2,0){\epsfxsize=18cm\epsfbox{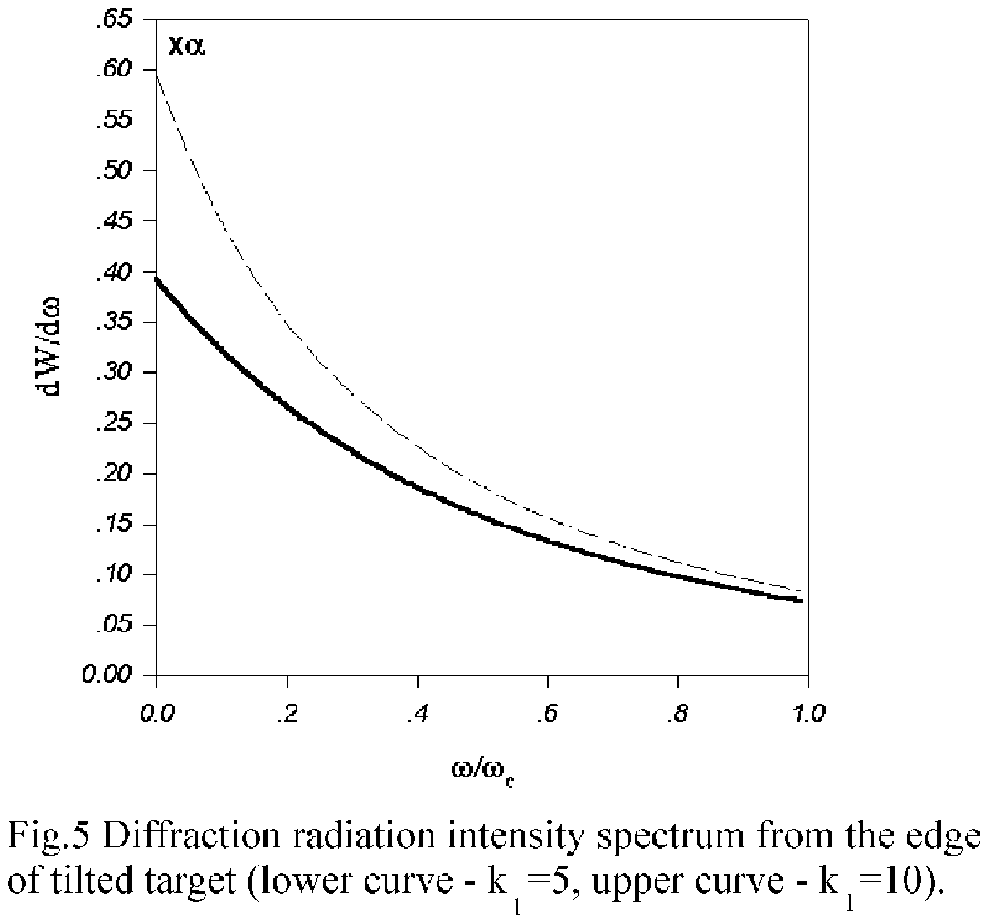}}
\put(-2,-4){\epsfxsize=18cm\epsfbox{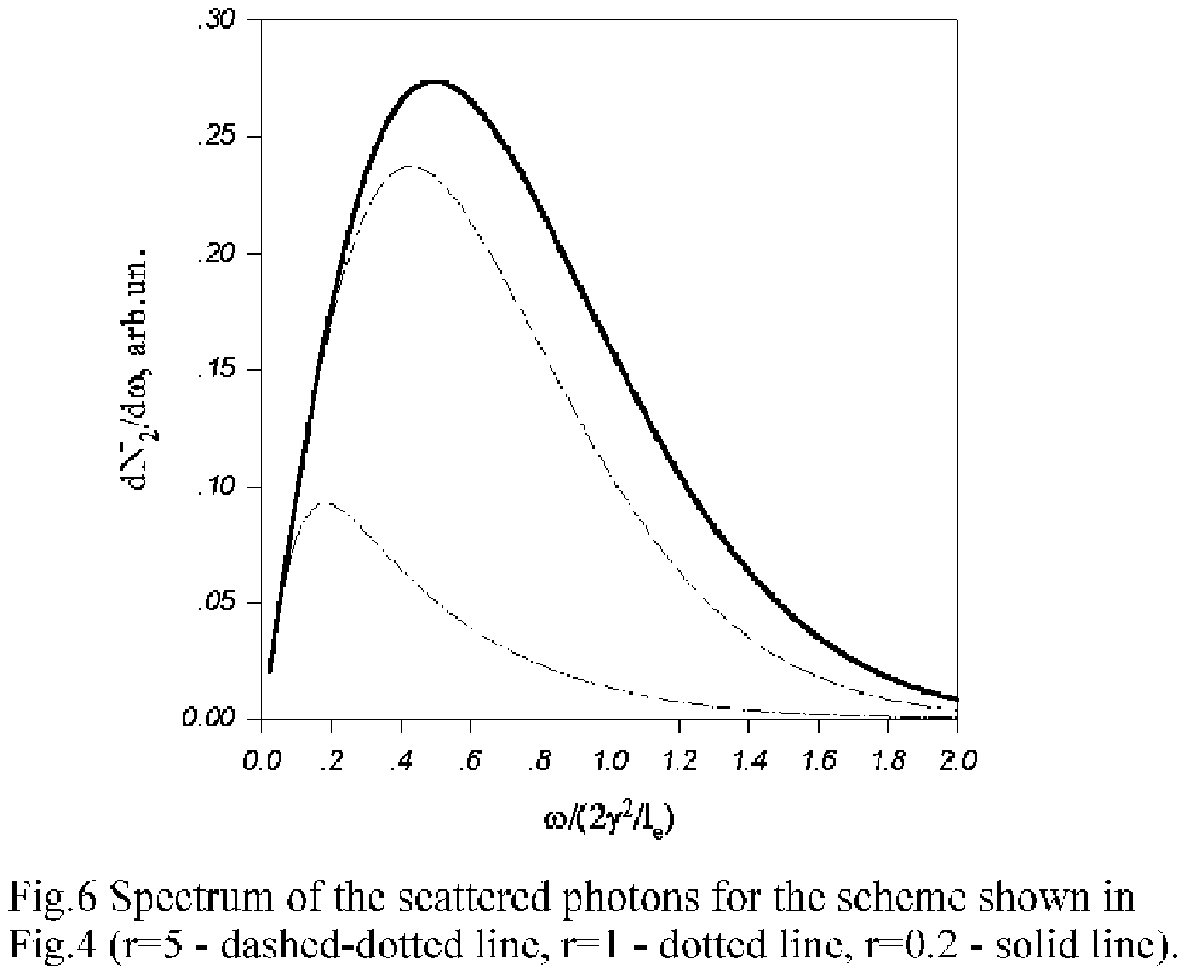}}
\end{picture}
\end{figure}

\end{document}